\documentclass{mpe_report}
\usepackage{psfig}
\usepackage{natbib}
\newcommand{\aaps}{A\&AS}
\newcommand{\aap}{A\&A}
\newcommand{\apj}{ApJ}
\newcommand{\apjs}{ApJS}
\newcommand{\araa}{ARA\&A}
\newcommand{\pasj}{PASJ}
\newcommand{\aj}{AJ}
\newcommand{\mnras}{MNRAS}
\newcommand{\ergcm}[1]{$\times 10^{#1}$ erg cm$^{-2}$ s$^{-1}$}

\newcommand{\ergs}[1]{$\times 10^{#1}$ erg s$^{-1}$}
\newcommand{\oergs}[1]{$10^{#1}$ erg s$^{-1}$}

\newcommand{\ohcm}[1]{$10^{#1}$ cm$^{-2}$}
\newcommand{\cts}[1]{$\times 10^{#1}$ cts s$^{-1}$}
\newcommand{\ct}{cts s$^{-1}$}
\newcommand{\Hone}{\ion{H}{I}}
\newcommand{\Halp}{H${\alpha}$}

\begin{document}
\title{X-ray source populations in the Magellanic Clouds}
\author{F. Haberl\inst{1} \and W. Pietsch\inst{1}}  
\institute{Max--Planck--Institut f\"ur extraterrestrische Physik,
 Giessenbachstra{\ss}e, 85740 Garching, Germany}
\maketitle

\begin{abstract}
  Early X-ray surveys of the Magellanic Clouds (MCs) were performed with the
  imaging instruments of the Einstein, ASCA and ROSAT satellites
  revealing discrete X-ray sources and large-scale diffuse emission.
  Large samples of supernova remnants, high and low mass X-ray binaries
  and super-soft X-ray sources could be studied in detail. Today, the major
  X-ray observatories XMM-Newton and Chandra with their advanced angular
  and spectral resolution and extended energy coverage are ideally
  suited for detailed population studies of the X-ray sources in these
  galaxies and to draw conclusions on our own Galaxy. We summarize
  our knowledge about the X-ray source populations in the MCs from past
  missions and present first results from systematic studies of the
  Small Magellanic Cloud (SMC) using the growing number of archival 
  XMM-Newton observations.
\end{abstract}

\section{Introduction}

The study of X-ray source populations and diffuse X-ray emission 
in nearby galaxies is of major importance in understanding the 
X-ray output of more distant galaxies as well as learning about
processes that occur on interstellar scales within our own Galaxy.
The MCs, satellites of the Milky Way, show different chemical 
compositions, are irregular in shape, and are heavily interacting 
with the Milky Way. This influences the process of star formation and the
study of stellar populations in the MCs is particularly rewarding. 
Their proximity makes the MCs the ideal galaxies for X-ray studies. 

Previous X-ray surveys of the MCs, which were performed with the imaging 
instruments of the Einstein, ASCA and ROSAT satellites, revealed discrete 
X-ray sources and large-scale diffuse emission. 
The early Einstein observations unveiled more than one hundred 
point-like sources in the Large Magellanic Cloud 
\citep[LMC;][]{1981ApJ...248..925L,1991ApJ...374..475W} and seventy
in the SMC \citep{1992ApJS...78..391W}. 
ASCA found more than 100 sources and detected coherent pulsations from 
17 sources in the SMC \citep{2003PASJ...55..161Y}. 
In particular the high sensitivity and the large field of view of the 
ROSAT PSPC provided the most comprehensive catalogues of discrete X-ray 
sources in the directions of the LMC
\citep[758 in an area of $\sim$59 square degrees; ][]{1999A&AS..139..277H} 
and the SMC
\citep[517 in an area of $\sim$18 square degrees; ][]{2000A&AS..142...41H}. 
Together with ROSAT HRI observations this yielded about 1000 and 550 
X-ray sources in the areas of the LMC and SMC, respectively \citep{2000A&AS..143..391S,2000A&AS..147...75S}.
A spectral analysis of the emission from the hot thin plasma in the 
interstellar medium (ISM) using ROSAT PSPC data of the MCs revealed
temperatures between 10$^6$ and 10$^7$ K 
\citep{2002A&A...392..103S}. 

Complementary deep surveys at other wavelengths were carried out by, e.g., 
{\it Galex} (UV) and {\it Spitzer} (IR), together with ground-based observations of 
\Halp, \Hone\ and radio continuum. In the optical/NIR bands the principal large 
scale digital surveys of the MCs are the micro-lensing surveys (MACHO and OGLE), 
the bright star survey of \citet{2002ApJS..141...81M}, 
the MCPS Magellanic Clouds Photometric Survey \citep{2002AJ....123..855Z} 
and the DENIS and 2MASS near infra-red surveys.
Combining the available information from the different wavelength bands allows 
to characterize the properties of different X-ray source classes like supernova 
remnants (SNRs), high and low mass X-ray binaries (HMXBs, LMXBs) and super-soft 
X-ray sources (SSSs) and differentiate them from background sources (mainly 
active galactic nuclei, AGN) or foreground stars. For the classification of 
sources X-ray spectral information (hardness ratios), X-ray to optical flux 
ratio, angular extent and flux variability can be used 
\citep[e.g., ][]{2000A&AS..142...41H,2003PASJ...55..161Y,2005MNRAS.362..879S,2007MNRAS.376..759M,2007arXiv0710.2232M}.
This has also been demonstrated for other Local Group galaxies like M31 \citep{2005A&A...434..483P}
and M33 \citep{2004A&A...426...11P}.

\begin{figure*}
\hbox{\psfig{figure=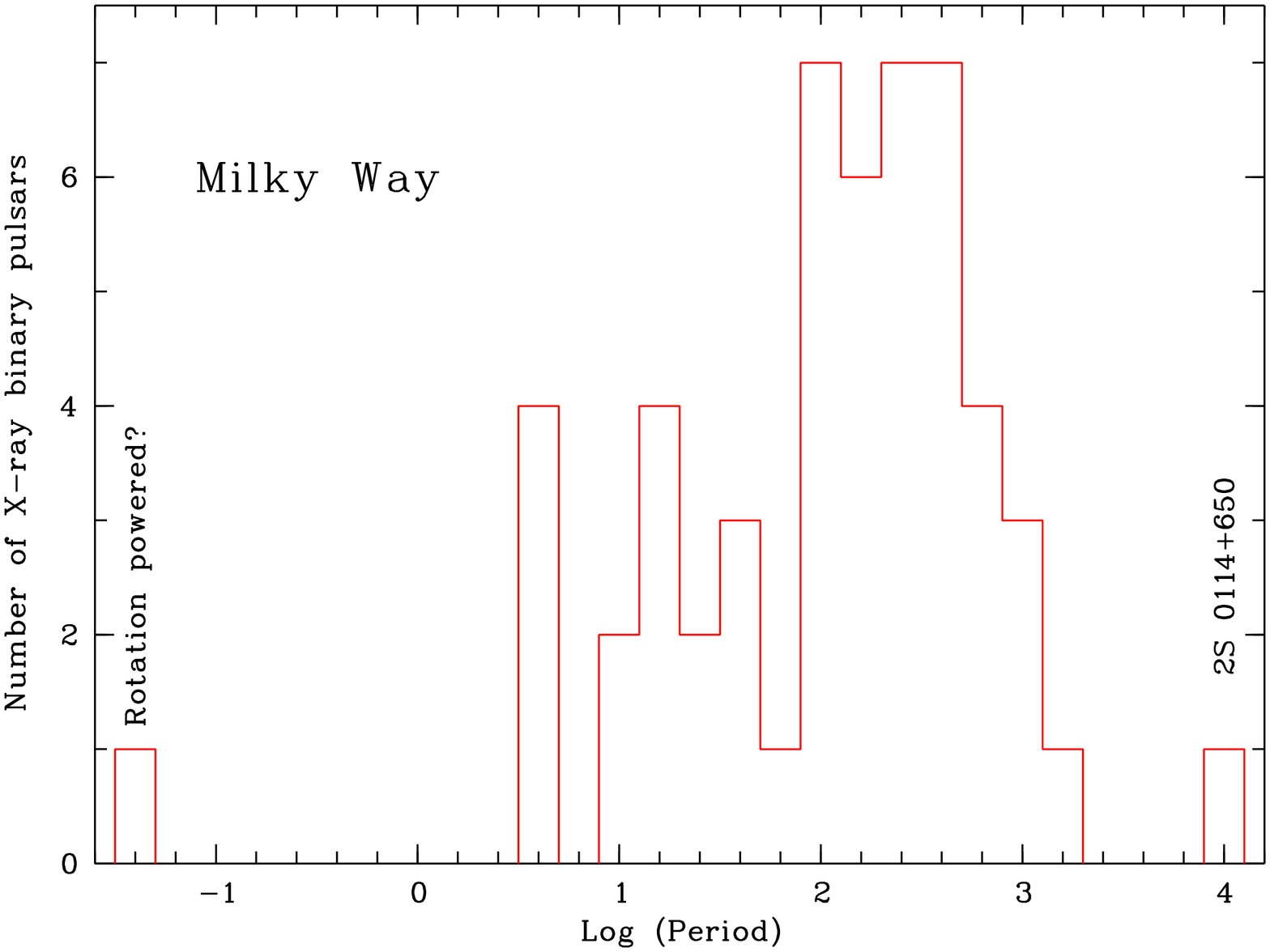,angle=-90,width=86mm}
      \hspace{6mm}
      \psfig{figure=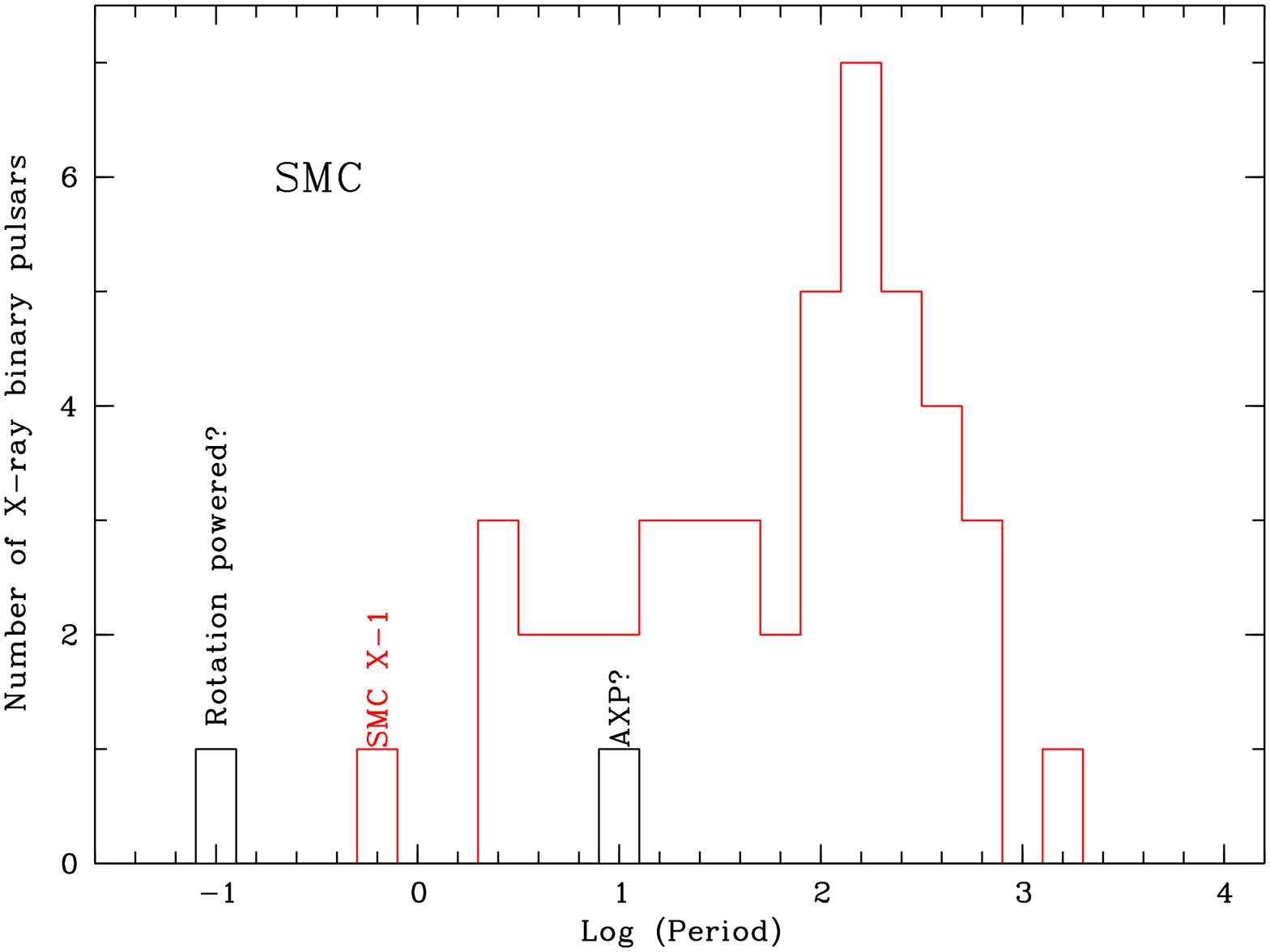,angle=-90,width=86mm}}
\caption{Comparison of the pulse period distribution of HMXB pulsars in the Milky Way
  and the SMC. Despite the large difference in galaxy mass, nearly as many HMXB pulsars 
  are known (status beginning of 2006) in the Milky Way (53) and the SMC (46).
  In both galaxies most of the pulsars are found with spin periods between 100 s and 
  1000 s. The fraction of pulsars with spin periods between 1 s and 100 s is slightly higher
  in the SMC as compared to the Milky Way.}
\label{pulsars}
\end{figure*}

\section{Source populations}

The SMC is very rich in HMXBs, which can be used as tracers of recent star 
formation in the galaxy. The evolutionary age of these early type binary
systems is low \citep{2003MNRAS.339..793G} 
and most of the Be-HMXBs in the SMC are found in regions with young 
(20$-$50 Myr) stellar populations 
\citep{2007AstL...33..437S}. 
About fifty HMXB pulsars are now known in the SMC together with many more 
candidates which exhibit similar X-ray properties, but yet without detected 
pulsations which reveal the spin period of a neutron star
\citep[e.g., ][]{2000A&A...359..573H,2004A&A...414..667H}.
Such a large number of HMXBs can be used for statistical studies, e.g. for a 
comparison with the HMXB population in the Milky Way (see Fig.~\ref{pulsars}).
In nearly all cases in the SMC the mass donor star is a Be star 
\citep{2005MNRAS.356..502C} 
with a circum-stellar disc 
\citep{2001A&A...377..161O}. 
When the binary orbit is wide and eccentric, the passage of the neutron star 
close to the disc results in X-ray outbursts and a transient behaviour of the 
Be-HMXBs. 
Many of the Be-HMXBs in the SMC were discovered during outburst (often 
exceeding \oergs{37} in X-ray luminosity), in particular with ASCA 
\citep{2003PASJ...55..161Y} 
and RXTE
\citep[for results from a regular monitoring of the SMC see ][]{2005ApJS..161...96L}.

The X-ray spectra of many HMXBs show, in addition to the typical power-law, a low-energy
emission component. The origin of this soft excess is not clear yet \citep{2004ApJ...614..881H}. 
The Be-HMXBs in the SMC are ideally suited to investigate the soft part of their spectra 
due to the low foreground absorption, in contrast to HMXBs in the Milky Way which are mainly 
found in the galactic plane suffering high absorption.
The best case to study the soft spectral component so far is probably the Be-HMXB
RX\,J0103.6$-$7201 (with 1323 s the pulsar with the longest period known in the SMC) which showed
during one XMM-Newton observation a highly absorbed power-law component and a completely
disentangled soft component. The soft component can be reproduced 
by thermal plasma emission with its luminosity strongly correlated with the total intrinsic 
source luminosity, suggesting that the same mechanism is responsible for the generation of 
the soft emission \citep{2005A&A...438..211H}.
 
Luminous super-soft X-ray sources were discovered with the Einstein observatory
(CAL\,83 and CAL\,87) and were established as a new class of X-ray binaries after the ROSAT discoveries 
of five new such objects in the LMC
\citep{1994A&A...288..538K,1996LNP...472Q.299G}.
SSSs exhibit very soft X-ray spectra (characteristic temperatures
kT of a few tens of eV) and 
show a variety of intensity variations on different time scales
(little variations, slow exponential decay over years, transient outbursts, off-states).
The most popular model for SSS involves nuclear burning on the 
surface of an accreting white dwarf (WD) which can explain the 
observed luminosities \citep{1992A&A...262...97V,1997ARA&A..35...69K}. 
The WDs in SSS indicate an older population consistent with their distribution
mainly in the outer parts of the MCs.
After the detection in the MCs, SSSs were
also found in other local group galaxies. Interestingly the majority
of SSSs in M31 was identified with optical novae \citep{2005A&A...442..879P} 
which enter a SSS state some time after optical outburst with onset and
duration of the SSS state varying strongly from source to source.
The MCs are sufficiently close to detect SSSs at low luminosities which 
allows us to address the question if permanently low-luminosity 
(too low for the high accretion rates inferred from the models) SSSs exist, or if 
they are highly variable which could point to unstable nuclear burning.
XMM-Newton observations of faint SSSs in MC fields show that this class
is composed of very different objects
\citep{2006A&A...452..431K,2006A&A...458..285K,2007ApJ...661.1105O,2008A&A...sub.....K}.
Symbiotic stars, central stars of planetary nebulae and even a Be star 
\citep{2006A&A...458..285K}
were identified as optical counterparts.
Be/WD systems are predicted to be more numerous than Be/neutron star 
binaries and the fact that we have so far not discovered any clear 
Be/WD case needs to be explained by binary evolution models. 

SNRs are a major source of matter feedback into the ISM.
Supernova explosions correlated in space and time generate super-bubbles (SBs) typically
hundreds of parsecs in extent. SNRs and SBs are among the prime drivers 
controlling the morphology and the evolution of the ISM. 
A synoptic study of thirteen SMC SNRs observed by XMM-Newton 
\citep{2004A&A...421.1031V}
showed a range of different morphological features from shell-like to more irregular
structures. A spectral analysis with single-temperature non-equilibrium ionization (NEI) and Sedov 
models revealed the different evolutionary phases of the remnants.

\section{XMM-Newton observations of the SMC}

Up to now, no complete X-ray survey of the SMC exists with the currently operating 
large observatories Chandra or XMM-Newton. A partial survey concentrating on the 
Eastern Wing \citep{2007MNRAS.376..759M}, was performed by Chandra.
These observations covered areas in the outer parts of the SMC, and the majority of
the 523 detected sources were identified as AGN, while the abundance of HMXBs (four detected 
pulsars) is low compared to the SMC Bar \citep{2007arXiv0710.2232M}.
XMM-Newton observations were dedicated to various individual targets in the SMC
resulting in a very inhomogeneous coverage. In a first attempt to derive the luminosity 
function of HMXBs in the SMC, \citet{2005MNRAS.362..879S} analyzed the (mainly EPIC-MOS) 
2$-$8 keV data of nine observations and compared the observed number of HMXBs with predictions 
based on star formation rate (SFR) estimates. Depending on the SFR indicators used,
the abundance of SMC HMXBs can be consistent with that of the Milky Way and other nearby galaxies
or as much as a factor of ten higher. 

\begin{figure*}
\centerline{\psfig{figure=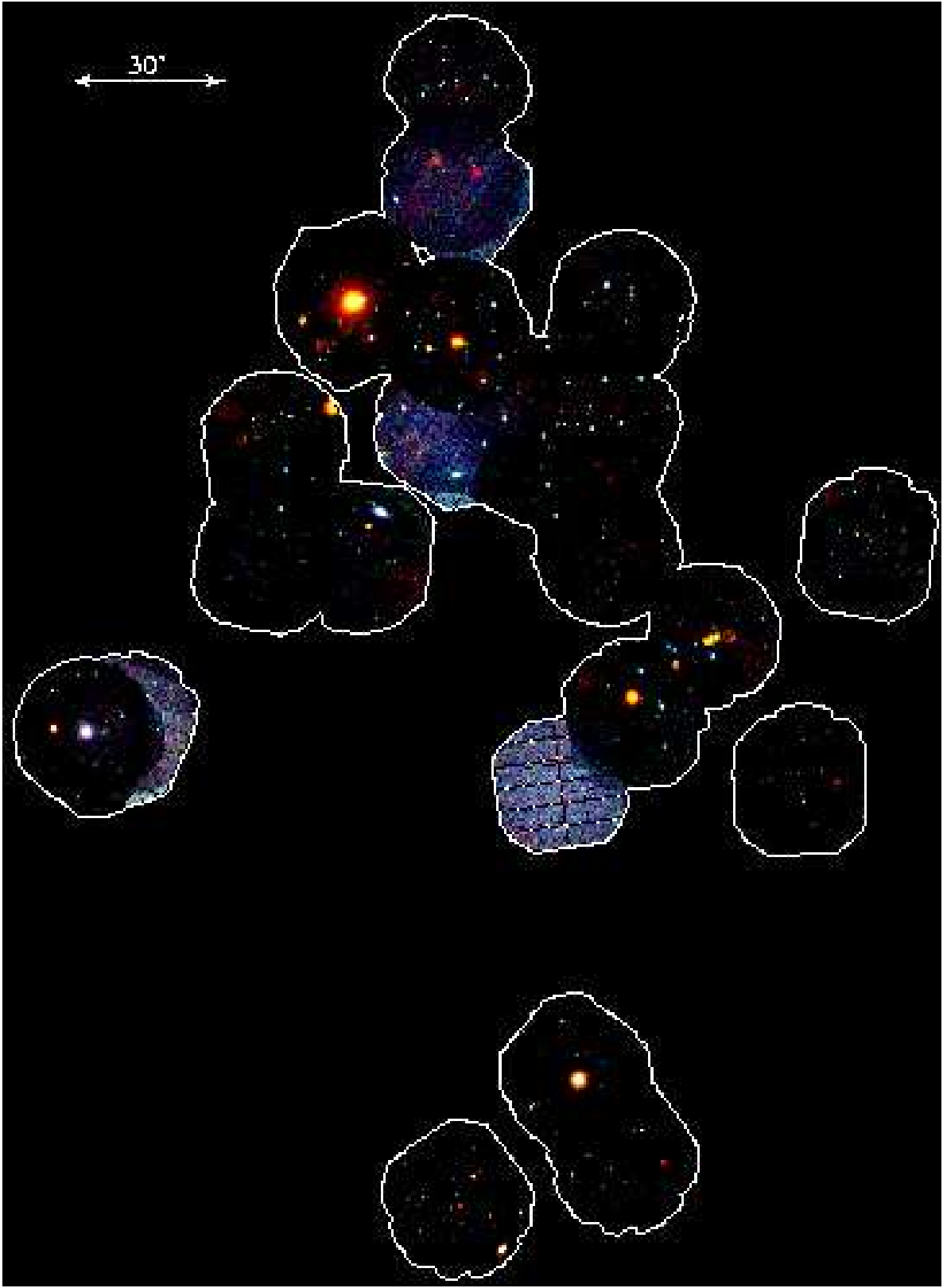,height=215mm,clip=} }
\caption{EPIC mosaic image of the SMC region obtained from the data of 35 individual observations. 
  The RGB colour image is composed of images from the three energy bands 0.2 $-$ 1.0 keV, 
  1.0 $-$ 2.0 keV and 2.0 $-$ 4.5 keV and from all three EPIC instruments. The individual images 
  are exposure corrected and out-of-time event subtracted (for EPIC-pn). The brightest source in 
  the north is the SNR 1E\,0102.2-7219. The observations used for this image
  accumulate to a maximum exposure of 210 ks (EPIC-pn) and 250 ks (EPIC-MOS) west of the SNR. 
  Typical exposures in other areas are 10$-$20 ks.}
\label{image}
\end{figure*}
                 
Meanwhile many more XMM-Newton observations are available in the direction of the SMC.
To obtain an interim XMM-Newton view of the SMC we systematically analyzed all
available EPIC data (the pn camera either in full frame (FF), extended full frame (eFF) 
or large window (LW) mode). This includes 
eight own proprietary observations from AO5 and AO6 together with all public 
archival data. After removing three observations which suffered from very high 
background, there remain 38 observations which we used for our analysis. 
Fourteen of them are calibration observations in the direction of 1E\,0102.2-7219 
which accumulate to a deep exposure in the north-eastern region. 
A mosaic RGB colour image is presented in Fig.~\ref{image} which is produced from 38 pointings
which mainly cover the Bar and the Eastern Wing of the SMC (three observations are pointed further
south around the RS CVn type variable CF Tuc).
The image includes data from four observations with slightly increased background 
which are not well suited for a clean image and the analysis of diffuse emission, but can 
still be used for the analysis of point sources. 
In the following we present a few examples of our first results with the emphasis on SNRs 
and HMXBs.

Three additional known SNRs -- SNR\,B0039$-$73.9 = HFPK\,530 
\citep{1998A&AS..127..119F,2000A&AS..142...41H,2007MNRAS.376.1793P}, 
SNR\,B0050$-$72.8 and  SNR\,B0058$-$71.8 \citep{1984ApJS...55..189M} -- 
are covered by new XMM-Newton observations. This allows to 
extend the investigations of SMC SNRs by \citet{2004A&A...421.1031V}. 
The three SNRs are located outside the main star forming regions where the majority of 
SNRs is found
\citep[north-east and south-west ends of the SMC Bar, see ][]{2001AJ....122..849D,2004A&A...421.1031V}.
All three show irregular X-ray morphologies with low surface brightness. An analysis
of the X-ray spectra with a single-temperature NEI model reveals relatively low 
temperatures around kT = 0.18 keV, suggesting that these are older remnants. 

\begin{figure*}
\hbox{\psfig{figure=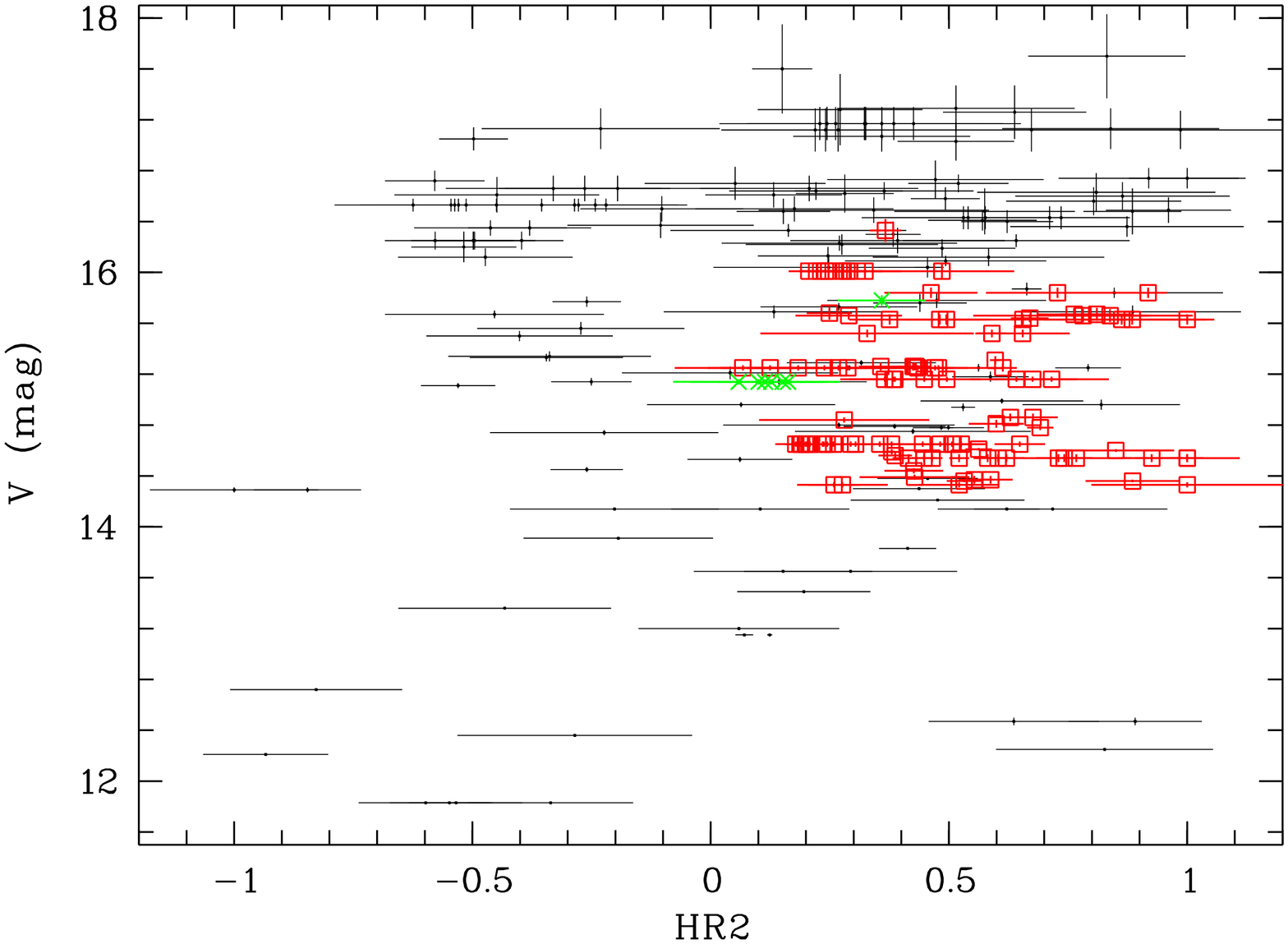,angle=-90,width=86mm}
      \hspace{4mm}
      \psfig{figure=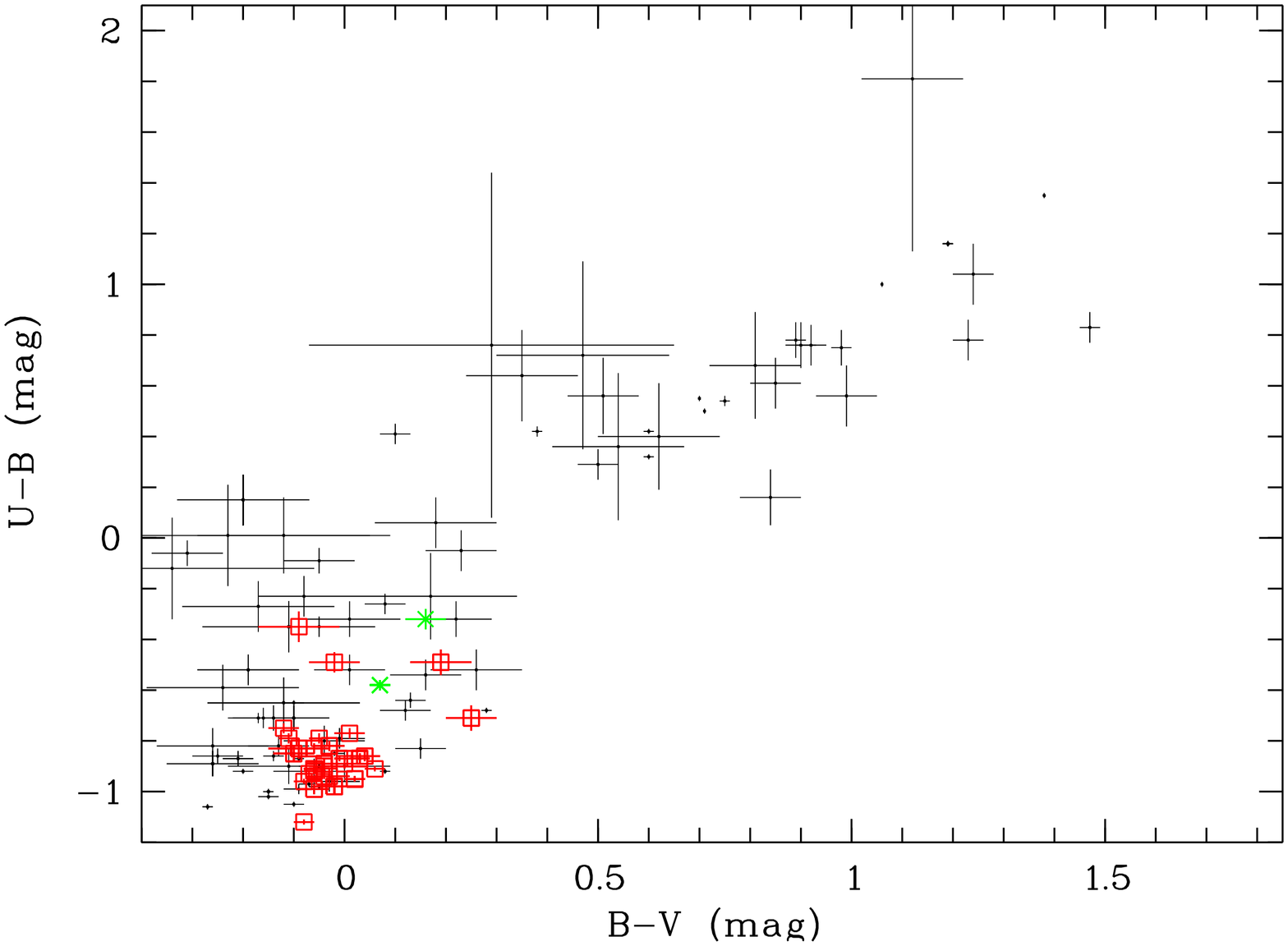,angle=-90,width=86mm}}
\caption{Correlation of the EPIC source detection list with the optical catalogue of 
  \citet{2002ApJS..141...81M}. The nearest star within 5\arcsec\ from the X-ray position is
  used as most likely optical counterpart. Multiple positional coincidences are rare because of the limiting magnitude
  of V = 18 of the optical catalogue.
  {\it Left:} V magnitude of the optical counterpart plotted versus hardness ratio 2 (restricted to sources
  with errors on HR2 smaller than 0.25). X-ray sources which are detected in different observations 
  may appear several times with different HR2 but the same counterpart (V magnitude). 
  Known HMXBs are marked with squares, AGN with x. Only very few AGN (two in the plots) 
  behind the observed SMC fields are as optically bright as HMXBs.
  {\it Right:} Comparison of B$-$V and U$-$B colour indices for the optical counterparts of the 
  X-ray sources. Known HMXBs in the SMC show typical V magnitudes of 14$-$16 and are closely concentrated
  in a small area in the optical colour-colour diagram.}
\label{correlations}
\end{figure*}
 
A new candidate SNR is proposed from its X-ray colours. It was already detected by 
\citet{2000A&AS..142...41H} in ROSAT PSPC data (source 334 in their catalogue
is listed with an extent of $\sim 18\arcsec$ and correlates with a radio source, consistent 
with emission from a SNR)
and following \citet{2004A&A...421.1031V} we name it HFPK\,334. 
Filipovic et al. (2008, in preparation) will present multi-wavelength morphological
studies and more detailed results from the X-ray spectral analysis of the three SNRs, mentioned 
above, and the new candidate.

The X-ray transient XTE\,J0103-728, discovered as 6.85 s pulsar by RXTE, was seen
in outburst at a 0.2$-$10 keV luminosity of 1.6\ergs{37} in October 2006. The EPIC data 
allowed us to accurately locate the source and to investigate its temporal and spectral 
behaviour. The identification with a Be star confirms XTE\,J0103-728 as Be/X-ray binary 
in the SMC \citep{2007ATel.1095....1H}. 
At least four new HMXB pulsars (Haberl et al. 2008, in preparation) are discovered. Their 
optical counterparts are identified with stars in the optical SMC surveys which show 
optical brightness and colours consistent with Be stars.

We performed a systematic source detection on the images obtained from the individual
observations (five energy bands for each of the three EPIC instruments). 
In total more than 1650 detections were obtained from the 38 analyzed observations.
This results in about 1060 individual sources.
The total detection list can be used for time variability studies for about 200 sources
which are detected more than once. 
The faintest sources have count rates as low as 1.3\cts{-3} (0.2$-$4.5 keV) in comparison 
to the brightest source (the SNR 1E\,0102.2-7219) with a pn count rate of 31 \ct.
Assuming a canonical HMXB spectrum (photon index 1.0, absorption column density \ohcm{21})
this translates into a typical flux limit (for detection) of 3.7\ergcm{-15} or a source luminosity of
1.8\ergs{33} for a distance of 60 kpc to the SMC.
With this limit, the subclass of persistent Be-HMXBs which shows typical luminosities of 
around \oergs{34} can easily be detected and also it should be possible to detect 
the first cataclysmic variables in the SMC when they are in bright state 
\citep[e.g. GK Per was observed with $\sim$\oergs{34} in outburst during EXOSAT observations;][]{1985MNRAS.212..917W}.

Source classifications using hardness ratios and optical information can be used to separate 
different source types and disentangle the SMC source populations from foreground 
stars and background AGN. In order to identify possible optical counterparts with high confidence,
systematic uncertainties in the EPIC X-ray positions need to be reduced. This work is currently
in progress by registering the positions of X-ray sources in individual pointings to an optical
coordinate frame using sources with well known counterparts. For this purpose the large number 
of HMXBs and the increasing number of quasars can be used that are known behind the SMC 
\citep[see ][ and references therein]{2003AJ....125....1G,2005A&A...442..495D}.
First results show that the systematic uncertainties in the EPIC positions can be reduced from 
typically 2-3\arcsec\ to about 1\arcsec.

As example to classify new HMXB candidates we correlated our EPIC detection list 
(in a preliminary way with unregistered X-ray coordinates) with the optical 
catalogue of \citet{2002ApJS..141...81M}. The X-ray spectra of HMXBs in the SMC 
are usually characterized by a hard power-law with a relatively narrow distribution of 
the photon index \citep{2004A&A...414..667H}. Therefore, the hardness ratio 2 (HR2) value 
is expected to cover a well defined range. Also their optical counterparts, in the SMC mainly Be stars,
show similar brightness and optical colours \citep{2005MNRAS.356..502C}. In this way we can
define areas in parameter space where sources of given nature (in this case HMXBs) are likely to 
be found. Figure~\ref{correlations} shows a comparison of the X-ray HR2 with V magnitude of their optical 
counterparts. The optical B$-$V and U$-$B colours \citep[from ][]{2002ApJS..141...81M} 
of the optical counterparts are also presented in Fig.~\ref{correlations}.
As expected the known HMXBs cover well defined locations in these diagrams and new candidate 
HMXBs can be selected 
\citep[for similar and additional selection criteria and methods see e.g. ][]{2005MNRAS.362..879S,2007arXiv0710.2232M}.

\section{Conclusions}

First results from systematic analyses of X-ray data of the SMC obtained with Chandra 
and XMM-Newton demonstrate the prospects for X-ray source population studies of the 
Magellanic Clouds, but also show the demand for a complete survey.
The full LMC is probably too large to be completely covered by the relatively small 
field of view of modern instruments in an acceptable number of observations. 
In the SMC up to now, particularly the outer regions with older stellar populations and the 
area of the Eastern Wing are only sparsely covered by XMM-Newton observations.
Full coverage is highly desired to derive complete source samples, 
probing areas of different stellar ages. Large-scale structures in the diffuse 
emission can only be investigated in a full survey. EPIC observations with relatively moderate
exposure are sensitive down to $\sim$2\ergs{33} for point sources in the SMC.
First results from studies of archival data demonstrate that one will be able to
characterize individual X-ray sources by their time variability and spectrum, 
if they are bright enough, and identify
and/or classify the sources using hardness ratios, long-term variability, source extent,
and information from other wavelengths.

\begin{acknowledgements}
We used data from XMM-Newton, an ESA Science Mission with instruments and contributions
directly funded by ESA Member states and the USA (NASA).
The XMM-Newton project is supported by the Bundesministerium f\"ur Wirtschaft und
Technologie/Deutsches Zentrum f\"ur Luft- und Raumfahrt (BMWI/DLR, FKZ 50 OX 0001)
and the Max-Planck Society.
\end{acknowledgements}
   


\end{document}